# Can models for long-term decarbonization policies guarantee security of power supply? A perspective from gas and power sector coupling

Andrea Antenucci[a], Pedro Crespo del Granado[b*], Blazhe Gjorgiev[a], Giovanni Sansavini[a*]

[a] Reliability and Risk Engineering Laboratory, Institute of Energy Technology, Department of Mechanical and Process Engineering, ETH Zurich, Leonhardstrasse 21, 8092 Zurich, Switzerland

[b] Dep. of Industrial Economics and Technology Management, NTNU, 7491, Trondheim, Norway

ABSTRACT

The assessment of adequacy and security of the energy system requires the detailed knowledge of physical and operational characteristics. In contrast, studies concerning energy transitions employ stylized models that oftentimes ignore the technical properties but have a lasting influence on long-term energy policies. This paper investigates the gap between energy system planning and operational models by linking these two perspectives: (1) a long-term investment model with low spatial resolution and high level of aggregation, and (2) a spatially resolved system security model that captures the interdependences between the backbone of the electric power sector, i.e., the electricity and the gas infrastructures. We assess EU decarbonization pathways of the electricity sector towards 2050 by integrating the investment decisions of the long-term planning model and the safety performance of the resulting system operations via the security assessment model. In a large RES deployment scenario, we investigate two flexibility options: gas power plants and cross-country transmission expansion. Using the integrated model, we analyze how the adequacy and security of supply under extreme short-term operational conditions impact the long-term planning of the energy system and the investment decision-making. We provide country specific recommendations for UK. Results indicate weaknesses in the gas-electricity system and suggest improvements on capacity allocation.

## 1. Introduction

Modelling and analyzing energy systems is becoming increasingly challenging due to the growing need to capture the interdependencies among various energy sectors and harmonizing different research viewpoints. The energy transition should not only encompass the analysis and proposition of long-term objectives and decarbonization alternatives but incorporate multi-layer energy system approaches that consider: 1) substitution effects among energy carriers, 2) complementariness in different models' geographical and temporal resolution assumptions, 3) a practical and reasonable level of technical detail and system security assessments, and 4) reciprocal effects and dependencies among energy sectors (buildings, transport, grids), and others. For example, in [1] authors highlight some modelling limitations on using long-term investment models as they lack technical details and aggregated geographical and temporal coverage compared to more technical models (e.g. power system approaches, see also [2]). In this regard, research on synergies and reciprocal effects (linkage or combination) between energy models has been referred as one of the next frontiers in energy system modelling [3]. Linking or combining models allows to harmonize and validate assumptions, exploit model capabilities (address model's weaknesses or stress strengths), provide more robust assessments, and challenge models boundaries by addressing cross-disciplinary research questions. To this end, in this paper, we investigate energy carrier integration by combining electricity and gas models along with a long-term investment model designed for energy transition analyses. The core objective is to use the technically detailed (country specific) electricity-gas model developed for system security assessment in order to provide feedback to the aggregated (long-term EU level) investment model. Based on this modelling framework, we analyze the adequacy and system security of the electricity infrastructure under different decarbonization pathways for the EU and the impacts to individual countries. For this analysis, we assume that gas-fired power plants (GFPP) will play an important role on balancing a large-scale deployment of wind and solar generation in Europe.

---

* Corresponding authors:
Email: sansavig@ethz.ch (G. Sansavini); pedro@ntnu.no (P. Crespo del Granado)

In this regard, security of supply[†] will face new challenges with the increasing share of renewable energy sources (RES). Planning for the security of supply faces additional challenges when the interdependencies between the electric and gas networks are considered. The interdependencies originate from the use of the GFPP to compensate for the volatile nature of the RES, where the former is supplied by the gas network. For example, strong interdependencies existing in electric and gas network operations noted by [5, 6], may lead to supply shortage to customers in both systems. This was the case in the US, after a cold weather event in 2011, when gas curtailments to GFPP and poor quality of gas supply accounted for 10% of production losses, i.e. 120 MWh [7]. Furthermore, such interdependencies are expected to be more prominent, as renewables become the largest source of power supply [8, 9]. The role of inter-connector capacities and of their expansion between neighboring countries in Europe has been investigated in [10] for the ability to mitigate the variability of integrated large-scale renewables in a cost-effective manner. Recently, studies claiming the feasibility of 100% renewable power systems have been critically reviewed [11], and, in fact, the case for feasibility is deemed inadequate for the formation of responsible policies. This has opened a stimulating debate in the domain of 100% renewable power systems about methodological and modelling choices [12]. In the face of the transition to a 100% RES supply scenario, GFPP are expected to balance the volatility of RES if no other supply flexible options are available (e.g., biomass or storage) [13, 14]. The interactions between gas and electric infrastructures occur via GFPP and electricity-driven compressors. These linking points couple operational dynamics that evolve on different time-scales and may increase the vulnerability of both infrastructures. The risk of disruption to the supply to customers due to interdependencies is connected to several factors, such as the characteristics and the amount of GFPP, the supply capability of the gas network and the spatial distribution of the off-takes, and the effects of market and contract agreements [15]. Depending on these factors, possible risks include the loss of gas supply to GFPP (e.g. excessive non-electric demand), the rupture of pipelines, compressors or limitations in imports, and the lack of electricity to electricity-driven compressor stations. These occurrences can possibly mutually affect each other and generate a cascade of failures, i.e. the sequence of one or more dependent component outages that are initiated by one or more common disturbances [16].

---

[†] The literature defines the security of supply as the capability of the electric system to withstand disturbances [4].

Several models describe the issues and benefits of interdependent gas and electric networks. The optimization of combined gas and electric systems, where a cost function is minimized to determine the generation level of generators and gas intakes in standard operations, is a common exploited approach. Alternatively, authors often combine the optimization of the electric system with the simulation of the gas infrastructure [17]. In short-term system planning studies, the detailed network operations are computed via physical-flow models. For the electric system modelling, both the AC [18, 19] and the DC [20, 21] power flow models are employed. For the gas network modelling, transient models are often preferred to steady state models, due to their ability to capture system dynamics despite a larger computation effort [22, 23]. In the risk related literature, few papers investigate the impact of faults and disruptive contingencies on the coupled operations of electric and gas networks. In [24, 25] a graph-based methods are exploited to assess the impact of removing random network nodes in both grids. In [18], a coupled steady-state hydraulic flow model and AC electric power flow models address the effects of random failures on the coupled operations. Chaudry at al. [26] analyze via Monte Carlo simulations the effect of uncertainties in wind production, failures of components and gas supply, on the coupled electric and gas systems in Great Britain. Saldarriaga and Salazar [27] exploit an optimization method based on master-slave decomposition to investigate the impact of liquefied natural gas installations on the mitigation of supply and transmission contingencies.

The mentioned papers provide valid approaches to study the short-term planning and detailed operations of combined gas and electric systems. Therefore, they have the capability to complement long-term planning models, whose broad viewpoint (highly aggregated representation of the energy system) does not allow to focus on the single power plant operation and, in general, on network technical constraints or security aspects. As pointed out by Welsch et al. [28], accounting for the short-term perspective in a long-term power system planning model can result in different power plant dispatch and capacity investments, thus avoiding limited and inconsistent policy recommendations.

To address these limitations among modelling approaches, this paper investigates the European energy transition by analyzing to what extent an energy system planning model guarantees the security of supply. The goal herein is to investigate the benefits of combining two models with different temporal and geographical resolutions, where one provides an investment plan for the energy infrastructure, while the other assesses the physical operations of the energy networks and identifies security of supply issues. In particular, this methodology combines the long-term perspective of the **E**uropean **M**odel for **P**ower system **I**nvestment with (high shares of) **R**enewable **E**nergy (EMPIRE model) [29, 30] with a detailed network operation analysis via the **N**exus **S**ecurity **M**odel (NSM) [31]. EMPIRE is a capacity expansion model for investments in generation and transmission expansion considering aggregated power system features (technology mix and cross-border capacity) of European countries. The NSM comprises two models, i.e. an electric network model and a gas network model. The electric network is modelled via *N*-1 secure unit commitment problem, while the gas system is represented via a one-dimensional transient flow model, which accounts for the dynamic of compressors, imports, and storages. Gas-fired power plants are considered as coupling points between the two infrastructures. The integrated gas-electricity models allow the representation of initial disruptive contingencies, such as line disconnections, power plant failures and compressors shutdowns, and evaluate the state of the system by computing the power flow on electric lines, the generator set points, the RES curtailments, the gas and the electric load shedding. Because of these features, the NSM performs adequacy and security analyses of the EMPIRE investment recommendations (i.e. results). The objective of this integrated modelling framework is to address the following research questions: (a) what is the impact of short-term operations, detailed geographical representation and consideration of extreme events on system security for long-term energy transition assessments? (b) Do long-term decarbonization outlooks properly account for adequacy and security of supply? Based on our analysis, the results demonstrate the capability of the EMPIRE model to find acceptable solutions for a given energy transition pathway of a country. Furthermore, potential weaknesses in the electric and gas networks are identified and recommendations are given.

The paper structure is organized as follows: Section 2 presents the EMPIRE and NSM models; Section 3 discuss the energy transition pathways for Europe, the obtained results from the EMPIRE model and the scope of the linkage with the NSM; Section 4 presents the results obtained by the NSM applied on a selected case study system; Section 5 discusses the results; Section 6 provides conclusions and reflections for future work.

## 2. Models

This section introduces the power system planning model (EMPIRE) designed to assess decarbonization pathways for the European power sector, and the NSM that is a unit commitment model capable of performing reliability assessment of the coupled power and gas systems. In NSM, we define system reliability as encompassing system adequacy and system security. Section 3.3 details the approach for integrating the NSM and the EMPIRE in the adequacy and security analyses.

*2.1. EMPIRE model: Investments in electricity generation and transmission*

EMPIRE is a European Model for Power system Investment with (high shares of) Renewable Energy [29, 30]. It is a capacity expansion model that determines investments in electricity transmission and generation. Its objective is to minimize system costs for the European power system by including investment and operational costs. It follows a commonly used framework in energy system models to represent strategic (long-term investments) and operational decisions (hourly scheduling) in a perfectly competitive market. With these capabilities, EMPIRE can assess decarbonization pathways by considering the interplay among low carbon technologies with different characteristics such as solar PV, wind energy, carbon-capture and storage (CCS) and nuclear power.

EMPIRE includes a portfolio of generation technologies, categorized as follows: thermal or conventional power plants (nuclear, fossil generation, CCS, biomass), intermittent power generation (wind, solar, run-of-the-river hydro) and storage generation (reservoir hydro, pumped hydro, and battery storage). All technologies have a maximum capacity on their power generation output. For example, the thermal generation have ramping limits and operational fuel costs. The intermittent power generation use predefined production profiles (e.g. wind and solar patterns) for each country. The data profiles assumptions are explained in [32, 33]. In particular, RES generation profiles have been computed via standard profiles provided by the software Renewables Ninja [34]. The data for the electric power and gas demands are taken from the SET-Nav project [35, 36]. The Storage generation has a limit on total output over a time interval, e.g., pump storage is represented with a charging unit (pump), discharging unit (generator) and an energy reservoir, all with their respective capacities. For each technology, EMPIRE utilizes their technical specifications and costs in line with standard modelling practices for energy systems models (see e.g. [33, 37]).

Aggregated cross-border interconnectors represent the electricity transmission infrastructure. Internal national grids are greatly simplified. Meaning that the country is a single node in the model and internal grid operations are not considered (this is the so-called copperplate assumption). Power flows in the network mimic a transportation model, i.e., loop-flows are not considered. Briefly, the EMPIRE model implements the following features.

- The objective function minimizes the net present value investment decisions for generation-transmission along with the hourly operational cost of balancing decisions. We also apply a carbon price and load shedding costs. Carbon price is exogenously taken from PRIMES scenarios data and from other sources [34, 38]. The load shedding cost are based on [39].

- Hourly supply-demand balance constraints. EMPIRE long time horizon covers 2015 to 2050. EMPIRE investment periods are every 5 years and representative weeks (with hourly intervals) for each season are used to determine the scheduling of operations. These are scaled up for each 5 year investment interval. See similar approach in [40]**.**

- Generators capacity constraints and ramping constraints.

- Representation of the energy balance of batteries and hydro pumped storage as well as losses incurred in the charging/discharging process.

- Transmission network flow constraints between countries.

- Countries energy mix restrictions and characteristics.

The geographical coverage of EMPIRE takes into consideration the European Union countries plus Switzerland and Norway, and some Balkan states. For each country, we have collected information on existing generation and transmission capacities. Technology costs, fuel prices and other parameters are taken from diverse publicly available sources. These include CAPEX and OPEX as well as learning curves. For a more detailed explanation on sources and inputs refer to [32, 33]. The EMPIRE model is a large-scale linear program that includes approximately 15 million variables and 22 million constraints. Depending on the scenario and technology choice, the model might take up to 5 hours to solve.

### 2.2. NSM: short-term interdependent electric and gas networks analyses

The NSM is developed for performing security and adequacy analysis on interdependent gas and electric infrastructures. As such, it comprises an optimization framework for the solution of the unit commitment problem, and a gas simulation tool for the description of the transient gas dynamic [31].

#### 2.2.1. Electric system model

Electric network operations are represented as a mixed-integer linear programming problem. Formally, the optimization minimizes the cost of operations ($GRc_g$), stat-up ($Sc_g^u$), and shut-down ($Sc_g^d$) of each generator $g \in G$, the value of lost load ($VOLL$), the curtailment of wind ($WC_w$) and solar power cost ($SC_k$) for each wind farm $w \in WF$ and solar farm $k \in SF$, and the curtailment of run of the river ($RoR$) power generation cost ($Rc_r$), for each $t$ in the time span $H$ and time granularity $\Delta t$ i.e.:

$$Min \sum_{t=1}^{H} \left[ \sum_{g \in G} \left( P_{gt} \cdot GRc_g \cdot \Delta t + \alpha_{gt} \cdot Sc_g^u + \beta_{gt} \cdot Sc_g^d \right) + \Delta t \right.$$
$$\cdot \left( LS_t \cdot VOLL \right.$$
$$+ \sum_{r \in RoR} \phi_{rt} \cdot Rc_r$$
$$\left. \left. + \sum_{w \in WF} WP_{wt}^C \cdot WC_w + \sum_{k \in SF} SP_{kt}^C \cdot SC_k \right) \right] \quad (1)$$

where $\alpha_{gt}$ and $\beta_{gt}$ are binary variables that assume value one when a power plant is started up or shut down, respectively, $P_{gt}$ is the power output of generator $g$, $LS_t$ is the amount of load shed, $\phi_{rt}$ is the curtailed power from the $RoR$ unit $r$, $WP_{wt}^C$ is the wind power curtailment at wind farm $w$ and $SP_{kt}^C$ is the solar power curtailment at solar farm $k$. Eq. (1) is constrained by the power balance at each electric bus, the line rated capacities, the minimum up- and down-time, the minimum stable generation, i.e. the minimum level of sustainable output at which a power generation unit can operate, the capacity and the ramp-rate of generators, the system reserve requirements, and the availability of wind and solar resources. Import and exports are provided as inputs by the EMPIRE model and are here considered in the node balances. In addition, GFPP fuel availability constraints are added upon minimum pressure violations in the gas network, as detailed in Section 0.

*2.2.2. Gas system model*

The transient dynamic of gas flow in the network is described by the mass flow, Eq. (2), and momentum equations, Eq. (3) [41]:

$$\frac{\partial m}{\partial x} + a \frac{\partial \rho}{\partial t} = 0 \quad (2)$$

$$\frac{\partial \rho}{\partial x} + m \cdot \frac{\lambda \cdot |\omega|}{2 \cdot d \cdot a \cdot c^2} + \rho \cdot \left( g \cdot \frac{\Delta h}{l \cdot c^2} + (1 + b \cdot \rho) \cdot \frac{\Delta \theta}{\theta \cdot l} \right) + \frac{1}{a \cdot c^2} \cdot \frac{\partial m}{\partial t} = 0 \quad (3)$$

where $m$ is the mass flow rate ($kg/s$), $a$ is the pipe cross section ($m^2$), $\rho$ is the density ($kg/m^3$), $g$ is the acceleration of gravity ($m/s^2$), $h$ is the height of the pipe element ($m$), $\omega$ is the speed of the flow ($m/s$), $\lambda$ is the coefficient of hydraulic resistance, $c$ is the speed of sound ($m/s$), $d$ is the pipe diameters ($m$), $\theta$ is the absolute temperature ($K$), $l$ is the length of a pipe ($m$), $b$ is the gas constant ($m^3/kg$). The solution of Eq. (2) and Eq. (3) is obtained by employing the implicit finite difference scheme with intermediate step proposed in [41], which is commonly exploited in academic and industrial applications.

The knowledge of the pressure profiles in the network allows the computation of the linepack, i.e. the amount of gas stored into the pipelines, which is performed for the entire system and individually for specific areas. The linepack is a measure of the flexibility of the system in compensating fluctuating demands, and it is here exploited for computing GFPP fuel availability constraints, as detailed in Section 0.

Compressor stations are modelled with a constant pressure ratio and a nominal capacity of 50 MW. Terminals, i.e. where gas imports take place, and gas storages are assumed to have constant injection profiles [17], which are proportional to their nominal capacity. Therefore, the fluctuating demands are compensated by linepack variations on a second-by-second base. At the end of one balancing period, i.e. one day, linepack is restored to its initial level. In case of maximum pressure violations, gas injections from storages and terminals in the violated zones are reallocated to other parts of the network that have small linepack values.

*2.2.3. Gas and electric model interactions*

The NSM considers GFPPs as coupling elements between gas and electric networks. The gas off-takes that derive from GFPP operations, i.e. the electric gas demands, are computed as:

$$P_{GFPP} = M \cdot HHV \cdot \eta \tag{4}$$

where $M$ is the off-take mass flow ($kg/s$), $HHV$ is the higher heating value of natural gas (J/kg) and $\eta$ is the overall GFPP efficiency, specific for each GFPP typology. The electric gas demand contributes with the non-electric gas demand, i.e. gas required by industries and households, to the total gas demand that has to be delivered via the gas network.

Excessive off-takes may, however, induce minimum pressure violations at the nodes of the gas network, as calculated via Eq. (2) and (3). To relive this pressure condition, gas customers with non-firm contract, such as GFPP, must be shed. Formally, the GFPP power output is limited via the formulation of a new constraint, which is derived from [31]:

$$\int_{T_0}^{T_0+T^*} \sum_{g \in G} \frac{P_{gt} - \hat{P}_{gt}}{\eta_g \cdot HHV} \cdot dt \leq -G^C \tag{5}$$

Where $T_0$ is the curtailment starting time, $T^*$ is the curtailment duration, $\tilde{G}$ are all GFPPs in the violated area, $G^C$ is the gas curtailment, $\eta_{\tilde{g}}$ is the GFPP efficiency and $\hat{}$ represents quantities computed by the last run of the optimization. The electric optimization is then newly performed with the addition of constraint Eq. (5). The iterations between the electric optimization and the gas analysis tool terminate when no more pressure violations are found, or when no more electric load can be shed.

## 3. Linking energy planning and security models to analyze the energy transition

*3.1. European pathways for the energy transition*

The investment decisions in electricity generation and transmission expansion made via the EMPIRE model are sensitive to assumptions on fuel price projections, infrastructure development (e.g. realization of CCS technologies or grid upgrades), technology costs and learning curves, carbon price and GHG emission targets. Moreover, assumptions on socio-economic developments are central in framing long-term scenarios. These assumptions are typically performed by defining pathways and storylines on how different societal and technological developments will affect the transition to a low-carbon society. According to the EU commission [42] and the SET-Plan [43], the main routes for the decarbonization of the EU energy system are energy efficiency, nuclear, renewables and CCS. The implementation of these decarbonization options, however, raises questions on the long-term impacts they have on the power system infrastructure, i.e., "What are their cost differentials?". As noted earlier, these long-term perspectives require the evaluation of the performance and the evolution of different energy mix portfolios towards 2050. To do so, in this paper, we define European pathways for the energy transition based on the assumption that cooperation between nations and geopolitical conditions might paint different scenes on achieving climate goals. Therefore, we define, implement and analyze these EU energy transition pathways defined in the context of the SET-Nav project (see [44]):

- ***a "national champions" pathway:*** this storyline assumes that national interests play a stronger guiding hand. EU countries seek to maximize their use of locally available resources, and pan-European infrastructure and integration projects face resistance. It assumes a focus on national preferences, entrenchment towards business as usual, continuation of traditional incumbents and national available solutions dominate the supply sector.
- ***a "directed vision" pathway:*** this storyline assumes a context of cross-border cooperation and integration. It suggests a path-dependent trajectory for the EU energy system which is directed by the Commission's vision for an ever-closer energy union. The EU together have shared expectations for the development of the energy infrastructure.

Both pathways aim to the same decarbonization goal, i.e. the EU carbon reduction target of 90% by 2050. As a reference year, the EMPIRE model uses 2010 and the carbon goal applies to the region as a whole. In terms of practical model implementation, we assume the following exogenous features in the evolution of the power system for both pathways: 1) the prospects of nuclear development will be very limited, 2) CCS technologies are not considered, and 3) the pathways have the same data for demand, fuel prices, technological development and carbon price. These scenario assumptions create the need for large RES investments in EMPIRE in order to fulfil a 90% carbon reduction target. Hence, for each individual pathway analysis, EMPIRE decisions on capacities for electricity transmission and generation produces different conditions on the need for investing in a certain energy mix and, hence, rely on flexibility options (storage, gas power, demand response and others). Based on the definition of the "national champions" pathway, we restrict cross-border grid expansion in EMPIRE. Such a situation leads to investing in country capacity options instead of relying on grid flexibility and shared capacity among European countries. In contrast, for the "directed vision" pathway, grid expansion is a major feature. In short, we analyze two pathways, namely, one pathway that prioritizes flexibility of supply based on gas power plants and national solutions (hydro, biomass, and others), and one pathway that relies on cross-border capacity. For both pathways, EMPIRE implementation and data are partially based on the PRIMES scenarios used by the European Commission. Both pathways use the EUCO 27 scenario for carbon price, demand, and other datasets [38].

*3.2. Long-term decarbonization scenarios*

Table 1 summarizes main results and key metrics for the pathways analyzed by the EMPIRE model. The capacity mix of the "directed vision" pathway is affected by transmission capacity expansion. High shares of RES are possible thanks to the flexibility offered by transmission expansion. By comparing it to the "national champions" pathway, we notice that the lack of transmission expansion triggers a stronger need for backup capacity from storage (Hydro or battery), gas and biomass plants. Also, demand response supports the integration of RES in the "directed vision" pathway. Because there are no investments in demand response and storage options in the "national champions" pathway, the RES share reported is lower. Moreover, wind power capacity (especially offshore wind) is large in the "national champions" pathway because solar generation has a lower share. This creates a high average cost of electricity (see Table 1).

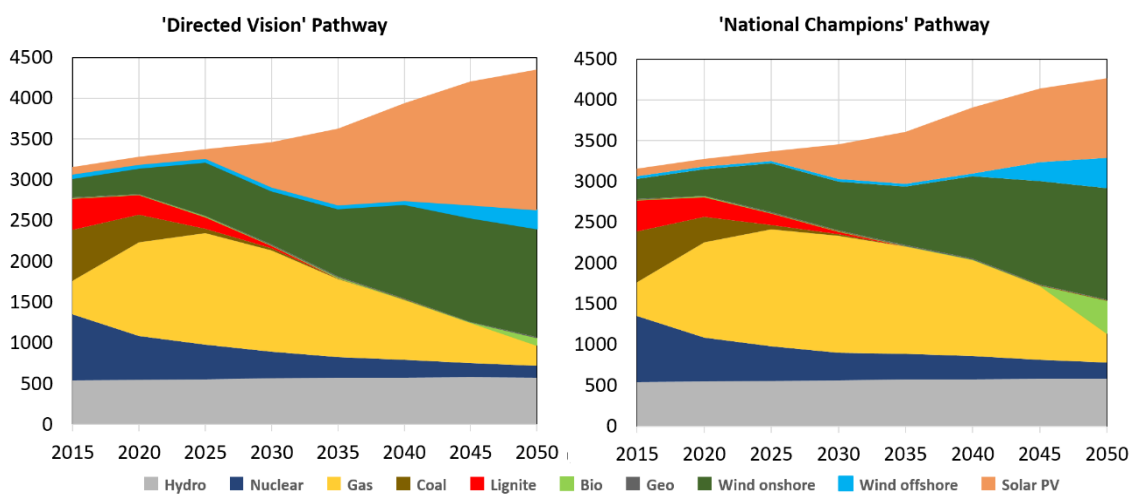

**Fig. 1.** Aggregated Generation (in TWh) for EU 28 (less Cyprus and Montenegro) plus Bosnia Herzegovina, Norway, Serbia and Switzerland.

A common trend in both pathways is the need of gas-based generation power as a transitional fuel to achieve emission reductions in 2050. Gas replaces coal and lignite plants from 2020 until 2030. The price of gas relative to coal price is determinant in this case as well as the high carbon price in the following decades. From 2030, gas declines progressively its annual generation in favor of solar PV and wind deployment. The gas abatement is much slower in the "national champions" pathway than its counterpart towards 2050 because there is less spatial flexibility as compared to "directed vision", i.e., limited expansion of storage and grid capacities.

**Table 1**
Summary of main results and key metrics (EU aggregated) for each pathway.

| Pathway | Year | Average electricity cost (€/MWh) | Generation adequacy without RES (%) | % of RES generation | % of Gas generation | % of storage generation & capacity | Emissions (MtCO2) | Curtailment (TWh) |
|---|---|---|---|---|---|---|---|---|
| **National Champions** | 2025 | 55.0 | 105.9 | 22.2 | 42.91 | 0.05 & 4.9 | 703.9 | 1.7 |
|  | 2035 | 66.0 | 97.8 | 38.6 | 36.43 | 0.17 & 3.3 | 434.6 | 22.6 |
|  | 2050 | 91.5 | 105.4 | 64.0 | 8.35 | 0.27 & 1.8 | 107.9 | 484.4 |
| **Directed vision** | 2025 | 54.6 | 101.0 | 24.2 | 40.61 | 0.05 & 4.9 | 682.0 | 1 |
|  | 2035 | 63.0 | 83.2 | 50.4 | 26.21 | 0.38 & 8.5 | 327.4 | 19.2 |
|  | 2050 | 74.2 | 75.8 | 77.0 | 5.90 | 1.42 & 16.1 | 78.6 | 406.8 |

In Table 1, "generation adequacy without RES" is the percentage ratio of the total conventional generation capacity and the year's peak demand and it quantifies the capability of covering peak demand using non-RES generation portfolio. This adequacy indicator shows that the "national champions" pathway results in a portfolio with more abundant conventional generation as compared to the "directed vision" pathway. For instance, the generation adequacy without RES for "national champions" does not decrease from 2025 to 2050 compared to "directed vision" case. This highlights that grid expansion decreases the need for conventional generation to cover peak demand. Similar trends are identified for the average electricity cost (74.2€/MWh versus 91.5€/MWh) and for the need of conventional capacity (gas generation 8.6% and 12.5%, respectively).

Overall, renewables are favored in the generation mix, but this requires flexibility (balancing) options to be deployed along with them. To this aim, gas-fired generation works as an intermediate solution for firm capacity, but it reduces its contribution to electricity production towards the end of the analysis horizon (2050). As shown in Fig. 1, RES generation increases, and gas generation decreases from 2025 to 2050 for both pathways. This is because of high carbon prices in 2030-2050, which call for other flexibility options ("greener" than gas power plants) to support RES integration. As a result, gas power plants are primarily used as base load units in 2025 but in 2050 they are mainly employed for balancing, and their utilization factor diminishes greatly in 2050 (i.e., from 65% in 2025 to 15% in 2050). Nonetheless, their flexibility and synergy with other technologies supports the increase in RES share. Key flexibility sources alternative to gas-fired generation emerge towards 2050 because they carry no emission costs (carbon price); this is the case of biomass in the "national champions" pathway. Also, storage charging / capacity plays a major role in 2050, storing 1.67% and 1.42% of converted energy in the "directed vision" pathway. Without expansion of the pumped storage hydro stations or batteries, as in "national champions", the annual stored energy would be 0.27%. Therefore, the different mix of flexibility options not only impacts RES deployment but also curtailments. For instance, the "national champions" pathway results in 80TWh additional curtailments in 2050 as compared to "direct vision" because the latter pathway has higher storage and transmission availability. Despite the available storage capacities in "directed vision", 406.8TWh from renewables cannot be absorbed (curtailed) by the system in 2050.

For additional insights on the results, Fig. 2 presents an overview of the generation mix (pie charts) for selected years and countries. The country color's intensity reflects the amount of import dependency (red) or the country's extra generation availability for exports (blue). In particular, the peak demand hour of the year is used to compute the percentage of imports or exports. More specifically, under the "directed vision" pathway, countries dependency increases greatly due to a stronger transmission capacity. In contrast, in "national champions" pathway, the countries are self-sufficient due to "national" flexibility options and slightly decreased RES share in the mix. In general, Fig. 2 shows that under the "directed vision" pathway transmission expansion strengths the collaboration between western and central Europe in 2050; conversely, under the "national champions" pathway each country takes a conservative approach prioritizing self-sufficiency. Energy mixes are comparable for both pathways in 2025, while they differ for 2050 as shown in Fig. 1.

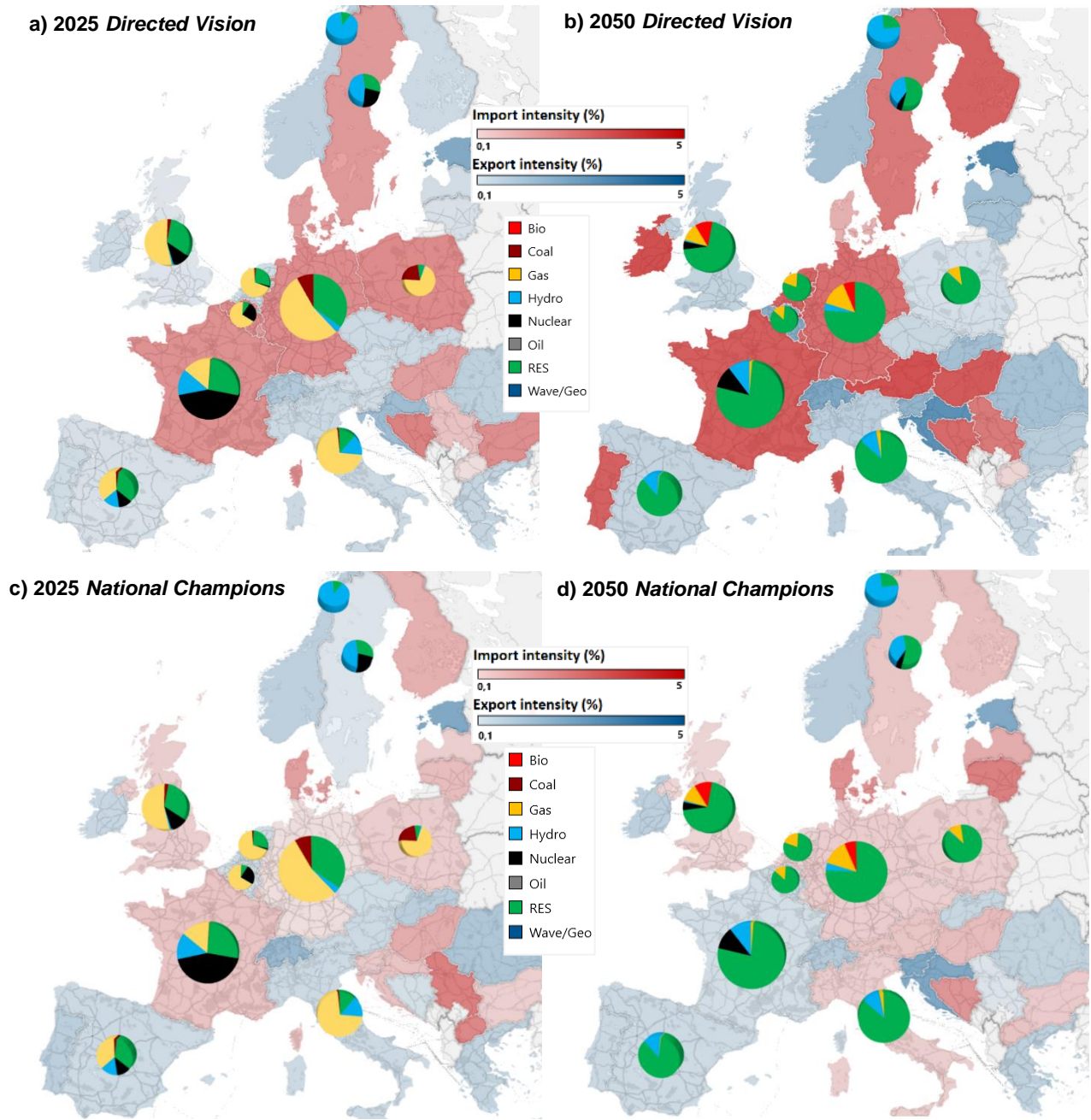

**Fig. 2.** Country generation profiles in 2025 and 2050. Red colored countries reflect the need for imports in the highest peak demand period (winter) while blue notes the country as exporter in that period.

*3.3. Combining models: EMPIRE – NSM synergies*

The EMPIRE results show the important role that RES have on achieving carbon targets for the EU power system. To do so, the model invests in biomass and gas power plants to support RES integration. The large RES deployment, however, does not consider where the wind farms will be allocated within the country. It also does not account for possible internal country grid bottlenecks and needed reinforcements, nor whether EMPIRE model decisions on cross-border grid expansions will create congestions in the interconnection nodes among countries. Also, a key aspect overlooked by long term expansion models such as EMPIRE, is the cross-sector effects to other energy carriers. If gas appears to be an important player to support RES, how does this affect the gas infrastructure within the country? Moreover, EMPIRE results might fall short on detail allocation of units in the country's transmission grid, where are these RES clusters located? Will these energy transition scenarios be feasible for the country's energy infrastructure? Nonetheless, the EMPIRE model provides a good overview on the policies needed to trigger investments in the power system at the EU level and countries.

The NSM described in Sect. 2.2, complements these limitations of the EMPIRE model. Due to the explicit modelling of the unit commitment, the representation of power flows, the consideration of component failures and of the interdependence with the gas network, the NSM provides a detailed spatial-temporal description of the conditions of the coupled system (gas-electricity nexus). Therefore, via the coupling of the two models, it is possible to address the technical issues that may arise in the gas and the electric systems. The long term planning by the EMPIRE model overlooks the operations within the countries energy system. Hence, the NSM can highlight vulnerabilities and indicate, for example, the need of building additional generation capacity or point out redundancies. This provides a direct feedback to the EMPIRE modelling approach. Furthermore, the NSM model results may identify bottlenecks in the gas and electricity transmission systems, thus, pointing to the necessity of strengthening the national network infrastructures, e.g. electric line or gas pipeline reinforcements and additional storage installations. The consideration of electric bottlenecks allows investigating the actual penetration of RES into the national system, since they may induce additional curtailments. For the same reason, accounting for the gas network operations is particularly valuable in the analysis of scenarios that entail large investments in gas-fired power plants as flexibility providers, given that gas-supply unavailability may compromise the full utilization of such plants.

The design of the EMPIRE – NSM models combination assumes the following interplay:

- The EMPIRE model provides as inputs to the NSM the following quantities, i.e., the generation capacity by sources at a national granularity, the hourly time series of the total electric load, the total solar and wind generation, and the hourly time series of electric imports and exports;
- The power exchange of cross-border interconnectors is implemented in the NSM model based on EMPIRE model results.
- Disaggregation of the EU-level model to country detailed gas-electricity models. To distribute the EMPIRE generation capacity within the country electric busses, we assume that the new capacity is built at the same locations of pre-existing power plants of the same type. This can be justified by the fact that many types of power plants require specific geographical and topological conditions to be operative, e.g. proximity to rivers or seas, difference in height and windy locations among others.
- Solar PV plants are uniformly distributed among electric busses, as it is detailed in Section 4.1. The hourly electric load is spatially distributed among electric busses by proportionally scaling a known load snapshot condition, as commonly done in similar studies [31, 45, 46]. Similarly, wind and solar plants contribute proportionally to their capacities, in order to match the respective exogenous time series.

## 4. Reliability analyses: UK case study

In order to display a detailed analysis on the gas-electricity interdependency, we focus on specific countrywide electric and gas networks and perform reliability and flexibility analyses for the United

Kingdom (UK) coupled electric and gas networks. The points of connection between the electric and the gas systems are the GFPPs, as shown in Fig. 3. The reduced electric grid consists of 29 electric buses and 99 overhead lines. Furthermore, the electric infrastructure is operated with a system spinning reserve requirement of 8 GW [47], which is the amount of back-up power capacity that needs to be available at any time for compensating load/supply uncertainties and possible faults of components. The reduced gas network consists of 9 terminals, 9 storage facilities, 69 pipelines and 21 compressor stations that work with a constant pressure ratio and a nominal power of 50 MW [31]. The pressure safety range in the gas network is 38 bar - 95 bar.

For the UK system, one iteration of the NSM model is completed in 1300 – 2500 s, depending on the analyzed pathway. Usually, a solution is obtained within 1 to 2 iterations. Calculations are carried out and parallelized on 24 cores (2.5-3.7 GHz) computer cluster [48].

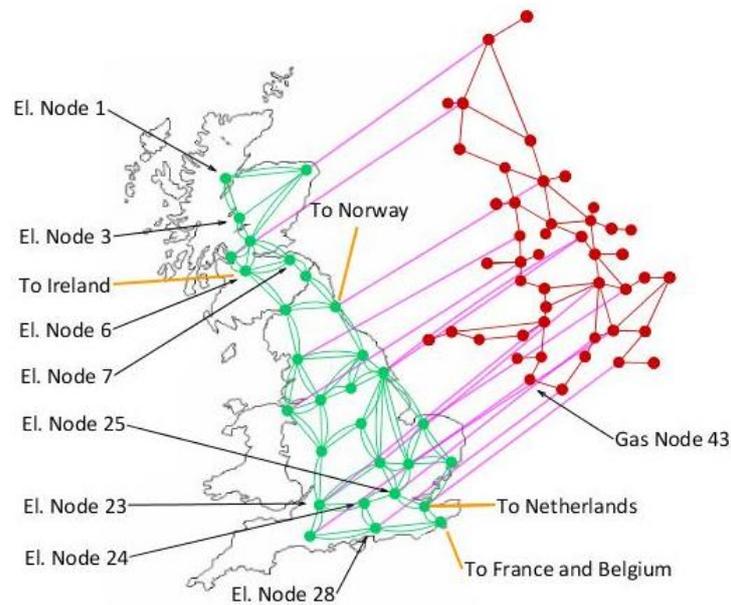

**Fig. 3.** The coupled electric (green) and gas (red) networks of UK [45].

### 4.1. Analyses set-up and selection of inputs

The "national champions" and "directed vision" pathways are assessed with the NSM to analyze the performance and evolution of different supply flexibility options towards 2050. The years 2025 and 2050 are analyzed as scenarios for each pathway. For each scenario, a 24-hour period that comprises the largest demand is considered. Furthermore, this 24-hour period[‡] accounts for the lowest RES generation output in the EMPIRE model. The non-electric peak gas demands are taken from [49] and are in correlation with the averaged gas demands given by PRIMES decarbonisation scenario (implemented in EMPIRE, see [38]), i.e. 341.4 mcm/d for year 2025 and 185.5 mcm/d for year 2050. The gas export to Ireland is 36.7 mcm/d in 2025 [49], and 19.9 mcm/d for 2050, which is obtained by scaling down the export value proportionally to the decreased gas demand of 2050. Overall, the reliability analyses of all scenario years (2025 and 2050) are performed for the extreme profiles for generation and demand assumed in the EMPIRE model investment analysis.

---

[‡] The EMPIRE model samples (snapshots) typical days per season (hourly country simulations) and also includes two days with a large peak demand and low RES profile events.

## 4.2. National champions – reliability analyses

In the "national champions" pathway, the EMPIRE model prioritizes the GFPPs as the flexibility providers in the power system with high penetration of RES.

### 4.2.1. System adequacy analyses

The goal of the adequacy analysis is to determine if there is enough capacity in the coupled electric and gas systems to supply the demand under improbable circumstances, i.e. one of the highest electric load demand, the lowest RES generation profile and the highest non-electric gas demand.

#### 4.2.1.1. Year 2025

The 2025 adequacy analysis (Fig. 4 (a)) shows that the electric system is able to supply demand in normal operating conditions.

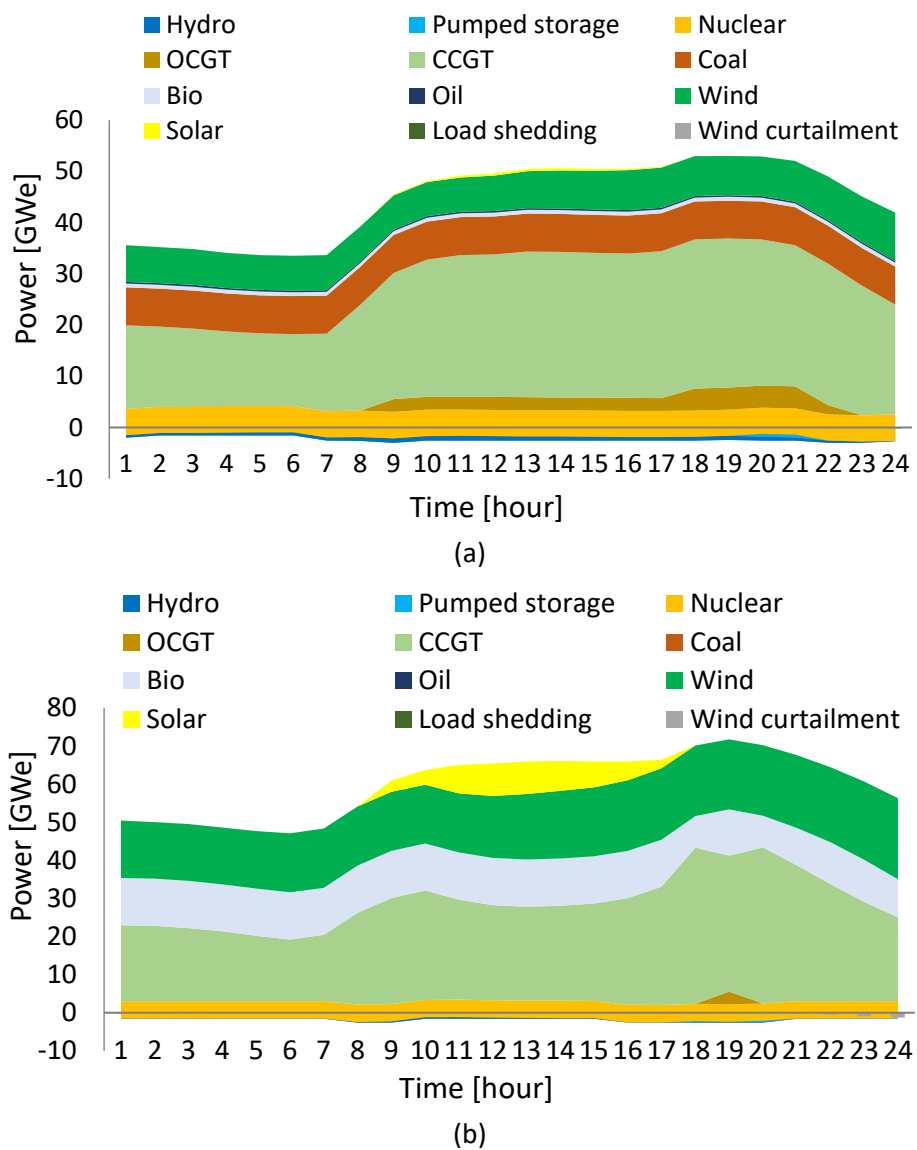

**Fig. 4.** Generation profiles for 2025 (a) and 2050 (b). The areas indicate the amount of power produced by different generation technologies at each hour. Negative values indicate power exports to neighboring countries.

However, minimum pressure violations occur in the gas network between hours 20 to 22 at the gas Node 43. The following gas curtailment to GFPP output amounts at 1.07 *$10^6$ $m^3$, which is handled by the electric system without supply interruption to electric customers. The 2025 generators and load shedding (LS) contributions to the total energy generation in percentage is given in Table 2, Row 1.

**Table 2**

Generators and load shedding contributions to the total energy generation in percentage.

| Scenario | Hydro | Pumped Storage (PS) | Nuclear | OCGT | CCGT | Coal | Bio/Lignite | Oil | Wind | Solar | Import | LS |
| --- | --- | --- | --- | --- | --- | --- | --- | --- | --- | --- | --- | --- |
| 2025 | 1.4 | 0.1 | 10.9 | 3.7 | 49.9 | 15.7 | 1.6 | 0.7 | 15.8 | 0.23 | 0.0 | 0.0 |
| 2050 | 0.3 | 0.0 | 7.3 | 0.2 | 42.1 | 0.0 | 18.9 | 0 | 27.6 | 3.6 | 0.0 | 0.0 |

The majority of the generation comes from CCGT, coal and wind power units. Results highlight that the capacity installed in the system and the existing gas transmission capability are sufficient to supply demand while satisfying a spinning reserve requirement of 8 GW.

*4.2.1.2. Year 2050*

Results in Fig. 4 (b) shows that the majority of the generation comes from wind, CCGT and bio/lignite generating units. An amount of 5 GWh of wind energy is curtailed ($WP^C$), with a maximum wind curtailment of 1.28 GW occurring during the last hour of the day. Curtailments occur in the Northern part of the network, at electric Nodes 1 and 3 due to line electric energy transfer limits, which may imply the need for additional capacity in the transmission system. It must be noted that no investments in the high-voltage transmission capacity within UK are considered in the "national champions" pathway.

*4.2.2. Linepack variations*

The linepack in the gas network varies in time due to the change of the gas demand level over the day. The nodal gas injections are constant, such that the gas network linepack is balanced every 24 hours as given in Figure 5.

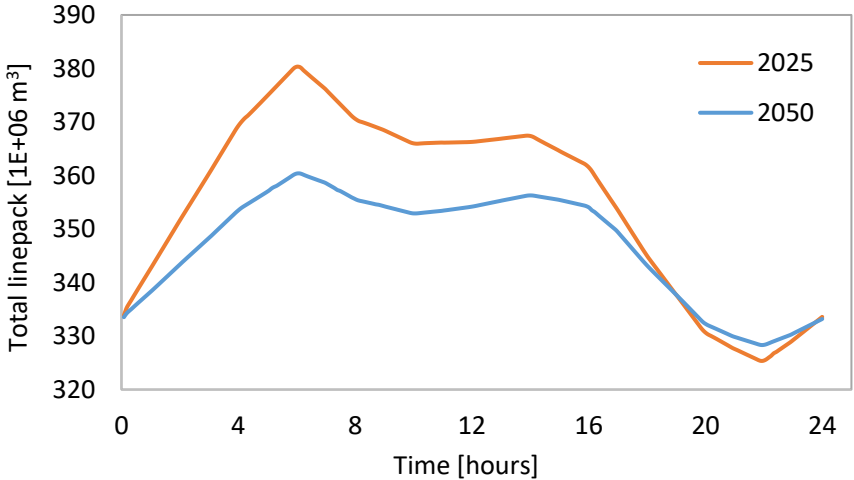

**Fig. 5.** Linepack variation for scenarios 2025 and 2050 in the "national champions" pathway.

Linepack variation decreases in the 2050 scenarios with respect to 2025 due to decreasing gas demand. In 2025 the maximum linepack variation is 46.5 mcm, while only 26.9 mcm in 2050. Results show that the gas system has the capacity to inject the required amount of gas even for large gas demand scenarios, such as in the 2025 scenario. However, such a large gas demand can cause minimum pressure violations, resulting in gas shedding (see Sect. 4.2.1.1).

*4.2.3. System security analyses*

The security analysis assesses if the coupled systems can withstand the loss of a single component. This type of assessment is known as *N*-1 security. The power system is normally operated in an *N*-1 secure state, and, therefore, we test whether such a security condition holds for the electric system designed by the EMPIRE recommendations. For the *N*-1 security analyses, we have selected 99 lines, 60 conventional power plants, 9 solar power plant clusters, 14 wind power plant clusters and 21 gas network compressors. One solar or wind cluster can contain more than one solar or wind power plants, respectively, all connected to the same transmission system electric bus. When a failure is simulated in one cluster, not all the plants belonging to that cluster fail, but only a portion of it is shut down. The NSM electric model employs aggregated conventional and RES generators, whose capacity may exceed several times the capacity of a real power plant. Therefore, the maximum loss of capacity for an aggregated generator is set at a maximum of 3960 GW (the size of the largest generating unit in UK [50]), and of 2000 GW for a wind farm. In the system security analysis, the system is not constrained with the required 8 GW spinning reserves, i.e. the unit commitment chooses the amount of generation to be deployed while minimizing load shedding after a contingency.

*4.2.3.1. Year 2025*

In year 2025, the failure of a CCGT connected to the electric Node 25 induces a small load shedding of 2.4 MWh ($2*10^{-4}$ % of total daily power energy demand). Furthermore, the loss of the overhead line connecting electric Nodes 23 and 24, lead to the loss of 0.1 GWh. No other contingency causes load shedding, proving that the coupled systems are able to avoid demand not served for the majority of considered failures.

Overall, results show that the pressure drop at gas network Node 43 is the major initiator of electric load and gas shedding in the electric and gas systems, respectively. Solutions to this problem may comprise the enhancement of the supply capability to Node 43 via the local installation of a new gas storage unit, the increased gas scheduling of neighboring storages or terminals and the construction of additional pipelines or compressors.

*4.2.3.2. Year 2050*

The *N*-1 security assessment for the 2050 scenario shows no pressure violations in the gas system. A small load shedding of 10 MWh follows the loss of the line that connects the electric Nodes 24 and 28, similarly to the loss of the wind farms at electric Nodes 6 and 7, which cause 14 MWh and 9 MWh of electric demand not served, respectively (less than $10^{-3}$ % of total daily power energy demand). Results show that the coupled systems can withstand the tested failures of single component.

*4.3. Directed vision – reliability analyses*

In the "directed vision" pathway, the EMPIRE model prioritizes the expansion of electric transmission interconnectors with the neighboring countries as the main source of flexibility.

*4.3.1. System adequacy analyses*

*4.3.1.1. Year 2025*

Fig. 6 (a) shows the daily generation profile for the "directed vision" pathway and scenario year 2025.

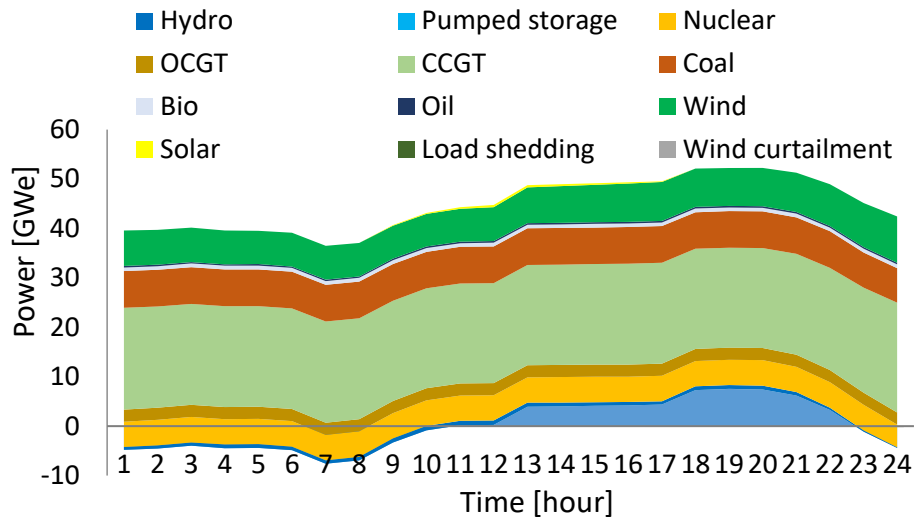

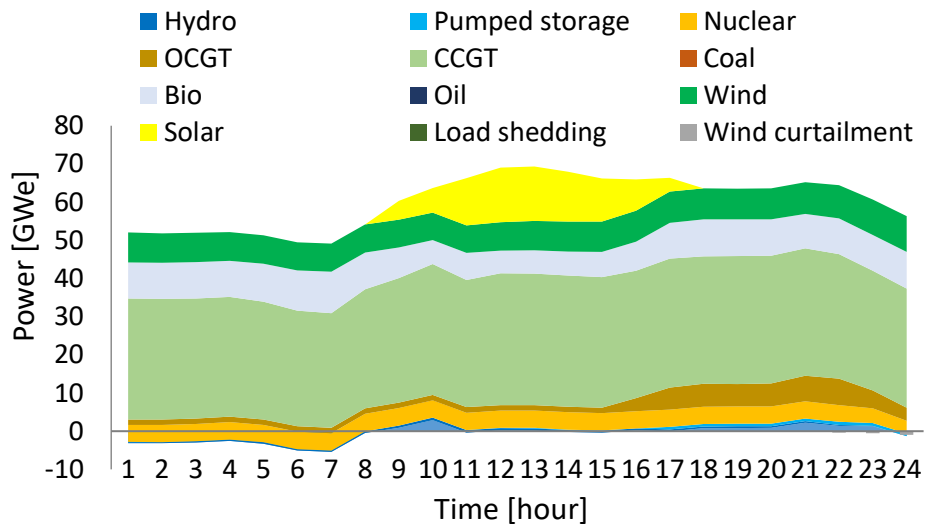

**Fig. 6.** Generation profiles for 2025 (a) and 2050 (b). The areas indicate the amount of power produced by different generation technologies at each hour. Negative values indicate power exports to neighboring countries.

The 2025 generators and load shedding contributions to the total energy generation in percentage is given in Table 3, Row 1.

**Table 3**

Generators and load shedding contribution to the total energy generation in percentage for the 2025 and the 2050 scenarios.

| Scenario | Hydro | PS | Nuclear | OCGT | CCGT | Coal | Bio/Lignite | Oil | Wind | Solar | Import | LS |
|---|---|---|---|---|---|---|---|---|---|---|---|---|
| 2025 | 1.5 | 0 | 10.9 | 5.2 | 43.7 | 15.7 | 1.6 | 0.7 | 15.8 | 0.2 | 4.7 | 0.0 |
| 2050 | 0.7 | 0.4 | 7.3 | 4.8 | 53.1 | 0 | 14.1 | 0 | 12.8 | 6.0 | 0.8 | 0.0 |

The majority of the generation comes from CCGT, wind, coal and nuclear generating units. The adequacy study for the scenario year 2025 shows that the coupled power and gas systems can handle the level of electric and gas demands without performing load shedding while maintaining the expected spinning reserve requirements of 8 GW. No pressure violations occur in the gas network.

*4.3.1.2. Year 2050*

Fig. 6 (b) shows the daily generation profile for the "directed vision" pathway and scenario year 2050. The 2050 generators and load shedding contributions to the total energy generation in percentage is given in Table 3, Row 2. Results shows that most of the generation comes from CCGT, bio/lignite, wind, nuclear and solar generating units. The adequacy study for the scenario year 2050 shows that the coupled power and gas systems can handle the level of electric and gas demands without performing load shedding while maintaining the expected reserve requirements. An amount of 4.1 GWh of wind energy curtailment occurs during the day, mainly at the electric Node 1 located in the Northern part of the network. The largest wind curtailment occurs at hour 24 and amounts at 978 MW. No pressure violations occur in the gas network. Similarly, as in the 2050 scenario year for the "national champions" pathway, the adequacy analyses show that, with the large penetration of RES in the system, additional national transmission capacity may be need.

*4.3.2. Linepack variations*

The linepack variation for 2025 and 2050 scenario years in the "directed vision" pathway are shown in Figure 7. In 2025, the maximum linepack variation is 35.7 mcm, while only 22.8 mcm in 2050. In both scenarios, fluctuations of minor entity occur in the "directed vision" with comparison to the "national champions" pathway. This reveals that the "national champions" pathway is characterized by a large imbalance between gas supply and gas demand during the day, compared to the "directed vision" pathway. The gas demand in these two pathways differs only for the gas supply to GFPPs, therefore, the GFPP fleet experiences larger ramp-up/ -down events in the "national champions" pathway than in the "directed vision" pathway (see Fig. 4 and Fig. 6).

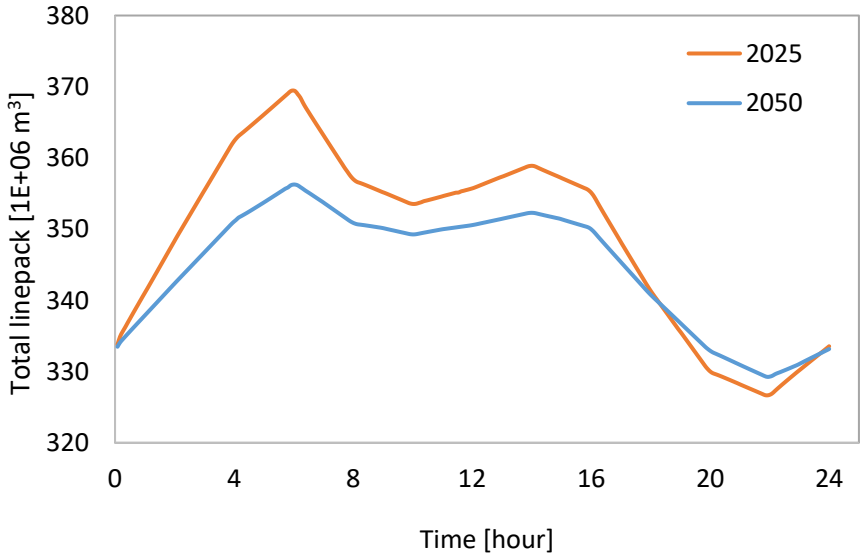

**Fig. 7.** Linepack variation for scenarios 2025 and 2050 in the "directed vision" pathway.

*4.3.3. System security analyses*

In the year 2025, the coupled power and gas systems have the capacity to avoid load shedding in all of the tested failures of single component events.

In the year 2050, the loss of several lines and power plants, i.e. 11% of the considered failures, leads to demand not served larger than zero. In particular, the loss of a CCGT connected to the electric Node 23 induces the largest load shedding of 16.5 GWh (1% of total daily power energy demand). Remarkably, the majority of load shedding events is caused indirectly by gas supply limitations to GFPP. In fact, this is the scenario year and pathway where OCGT and CCGT are utilized to a great extent. It is worth noting that the imports in the system are based on the EMPIRE calculations, which take into account the interconnectors in all Europe, and, thus, the NSM does not perform any import scheduling. Therefore, some load shedding can be eventually avoided if the constraints on the imported power scheduled via the EMPIRE model are loosened in the NSM model calculations.

## 5. Discussion: Experience and insights on combining models

EMPIRE endogenous investment decisions in generation and transmission provide a long-term outlook on the energy mix and infrastructure necessary for decarbonizing the power system. At the EU level, the EMPIRE model envisions the 2050 transformation of the EU power system towards a 70% renewable based generation system with gas power plants, biomass and storage as the key technologies to accommodate RES imbalances. But more importantly, EMPIRE proposes upgrades on cross border capacity to achieve an overall decreased cost in electricity prices compared to focusing on investing in country's individual technology mix. The "national champions" and "directed vision" contrasting pathways show the importance of promoting grid capacity investments and the need to incentivize flexible balancing technologies (e.g. capacity markets for gas power plants). The EMPIRE consideration of the net present value of investments under a long perspective and for a large geographical area (EU 28 plus) provides valuable and unique information which will be impractical to implement and obtain under the NSM approach due to computational challenges. The implementation of EMPIRE output as input to NSM demonstrated the advantages of combining the modelling approaches.

The "national champions" pathway in 2025 and the "directed vision" pathway in 2050 show that the large deployment of GFFPs during time of peak gas demand from the non-electric consumers can cause pressure violations in the gas networks ultimately resulting in electric load shedding. In particular, for the "directed vision" pathway in 2050, more investments in power plants are advisable, since the loss of generators cause supply interruption to customers. Moreover, the 2050 analyses show that with the increase of the RES generation in the system, up to 5 GWh of wind curtailment occurs. Therefore, additional transmission capacity within the UK electric network may be needed. Note that these assessments are made for the lowest RES generation profiles (representative day or week) assumed by the EMPIRE model. Therefore, it can be expected that the RES curtailments will be higher under different (more typical) RES profiles. Furthermore, in the "directed vision" pathway for the scenario year 2050, 11% of considered contingencies induce a load shedding of maximum 16.5 GWh. These observations confirm the aforementioned conclusion on the necessity of additional investments in the internal electric grid or the need of a combination of additional storage and grid reinforcements. Such investments can prevent the occurrence of operational issues and their associated costs, which are not considered by the EMPIRE model. For the EMPIRE model, this implies that additional modelling considerations and assumptions should be included or revised in its methodological framework (e.g. aggregation). Overall, the mostly positive NSM feedback indicates that EMPIRE's approach on sampling an extreme day (high peak demand and low RES) greatly influences the identification of a reliable energy generation mix. This is also in part due to EMPIRE is a multi-horizon stochastic program that represents short-term uncertainty (scenarios) which is not typical compared to other similar models.

## 6. Conclusions

In this paper, we analyze the energy transition of the power system by combining a long-term model for investments in electricity generation and transmission, with a joint physical gas and electric system models that account for the short-term operations and topology features of national grids.

The NSM analyses of the coupled UK power and gas grid, performed using extreme profiles of generation and demand, show that the long-term power system investments derived by the EMPIRE model can provide a satisfactory level of system adequacy. The "national champions" and "directed vision" pathways are adequate to handle the level of electric and gas demands without performing load shedding. The security of apply analyses of the "national champions" pathway show no significant concerns, while the "directed vision" pathway, for the scenario year 2050, results in load shedding up to 1% of total demand. In the scenarios with high penetration of renewables, a curtailment in wind generation due to grid congestions is pointing out to the necessity of grid investment into the internal electrical grid and additional storage. Overall, the combination of the EMPIRE and NSM results in a comprehensive assessment valuable for the planning of the future power systems with high RES shares.

Future research on methodologies that combine short- and long-term models should also consider the following points: (1) Long term investment models should consider or assume country local grid investments associated with their representation (and costs) of the technology mix; (2) Expand the number of countries under analysis since extreme conditions (Low RES scenario and high peak demand) in one country might influence neighboring countries system security and adequacy analyses; (3) Further study the curtailments reported by EMPIRE. For example, a stronger coupling and interaction among energy carriers (e.g. gas-electricity-heat) might provide a different perspective on actual curtailments. Also, power-to-X technologies should be considered in the EMPIRE portfolio mix as an alternative to handle RES surplus. Especially power-to-gas will be an attractive alternative option for storage if it's CAPEX and OPEX is competitive compared other technologies.

## Acknowledgements

This project has received funding from the European Union's Horizon 2020 research and innovation program / from the European Research Council under the Grant Agreement No 691843 (SET-Nav). The authors acknowledge the Swiss Competence Center for Energy Research - Future Swiss Electrical Infrastructure (SCCER-FURIES), for their financial and technical support to the research activity presented in this paper. Furthermore, the authors would like to explicitly acknowledge colleagues of the SET-Nav project: Sara Lumbreras, Andres Ramos, Luis Olmos, Quentin Ploussard, Christian Skar and Hector Maranon-Ledesma for their valuable and active participation in the SET-Nav report [44] to which this work is inspired.